# The learner who does not learn: when optimizing a pedagogical metric degrades LLM tutoring[1]

Daniel Dominguez Figaredo[a], Rafael Fernández De la Cruz[a]

[a] UNED, Madrid, Spain

ddominguez@edu.uned.es, rafael.fz@gmail.com

October 2026

**Abstract**

It is assumed that a natural way to improve the pedagogical quality of large language model tutors is to define a metric of instructional performance and fine-tune the model against it. To test this strategy, we designed a metric of pedagogical adaptivity that scores each instructional decision in a learning sequence against the conditions of the learning situation, which is the standard used for automated pedagogical scoring. We audited a frontier tutor across 2,000 learner scenarios, corrected its weakest cases by fine-tuning an open-weights proxy, and asked 31 trained educators to rate the pedagogical alignment of the outputs blind, before and after correction. The metric increased from +0.05 to +0.42 for the corrected cases, while the expert ratings decreased from 4.46 to 3.03, with the unmodified controls remaining unchanged and a base-proxy control ruling out the change of model. The tutor performed worse because any metric that scores decisions independently and averages them is maximized by repeating the single best decision, and the fine-tuned model collapsed to that exact optimum in every case, in and out of sample. Educators identified the repetition, which such metrics cannot represent, and preserving the learner's trajectory in the score reduced, but did not reverse, the metric's verdict. Weight analysis traced the correction to the model's output projection, where it had memorized its training strings rather than learned to adapt. We conclude that measurement validity does not imply optimization validity, and we derive design principles for benchmarks that assess or train AI-tutor instruction.

[1] This work was supported by the Tinker Research Program funded by Thinking Machines Lab. The authors have no relevant financial or non-financial interests to disclose. Data and code: https://github.com/MrEricPraline/pedagogical_adaptivity_in_LLM_tutors Corresponding author: ddominguez@edu.uned.es

## 1. Introduction

As large language models take on the role of tutor, the question of whether they reason pedagogically rather than merely produce fluent and plausible-looking instruction becomes consequential. AI tutoring is being adopted on the promise that it can deliver the individualized instruction that Bloom's (1984) two-sigma studies established as the benchmark (Chatterji et al., 2025; Bent & Handa, 2025), and recent accounts emphasize personalization as a key aspect of AI's role in education (Plass & Froehlich, 2025). However, several capacities that distinguish competent tutoring from content delivery remain largely unaudited. This study examines pedagogical adaptivity, the capacity on which claims of personalization rely most directly: the ability to adjust instructional decisions to the conditions that determine whether instruction will work for a given learner in a given situation (Cronbach & Snow, 1977; Koedinger et al., 2013). Leaving adaptivity unexamined matters asymmetrically. While educators can spot a content error at a glance, pedagogical misalignment remains invisible without systematic analysis. An instructional sequence may appear well-formed at every step yet still fail to meet the learner's needs, and a model may articulate sound principles while selecting activities that contradict them, a phenomenon described as "pedagogical drift" (Domínguez et al., 2026).

Current evaluations of LLM tutors present at least two limitations. First, they treat tutoring as a set of separate competencies, such as hints, corrections, and scaffolding, rather than as an adaptation to the conditions that define a learner's needs (Macina et al., 2025; Shi et al., 2026; Maurya et al., 2025), which means that a model can perform each move correctly without ever adapting its approach. Second, they only observe behavior from the outside, leaving the model's internal computation unexamined. This second limitation cannot be resolved by querying the model for analyzing its outputs and reflections. Chain-of-thought explanations often fail to reflect the computations that drive outputs (Barez et al., 2025; Chen et al., 2025; Lanham et al., 2023). Consequently, if an AI tutor's stated rationale does not correspond to its decisions, behavioral evaluation alone cannot determine if it reasons pedagogically. However, interpretability methods can determine not only if corrective fine-tuning changes behavior, but also where and how it does so (Burns et al., 2024; Cywinski et al., 2025; Xue & Mirzasoleiman, 2026; Shuttleworth et al., 2025). Therefore, this study combines a metric that is applied symmetrically to a model's decisions and educators' judgments of those decisions, resulting in an intervention that deliberately manipulates the decisions and provides a mechanistic account of the changes produced.

When pedagogical misalignment is detected, a natural response is to correct it by defining a metric of the missing quality and fine-tuning the model against that metric. Whether this improves instruction depends on how faithfully the metric captures the intended quality. Since pedagogical quality ultimately involves expert judgment, and automated and human evaluations of educational outputs tend to differ (Pal Chowdhury et al., 2025; Maurya et al., 2025; Szymanski et al., 2025), we tested this strategy by designing a metric of pedagogical adaptivity in the form used by current rubrics: it scores each of a tutor's instructional choices based on the conditions of the learner's situation, which are taken from a case narrative and remain constant throughout the activity sequence. This standard simplification is how learning theories are made measurable, and the study examines what this entails when the metric becomes a training target. We applied the metric to audit an AI tutor, corrected the tutor's lowest-scoring decisions through fine-tuning, asked trained educators to rate the outputs before and after correction, and examined the weights to see how the correction had changed the model.

The study is organized into two parts, both of which are centered around this instrument. The Pedagogical Adaptivity Index (PAI) scores an AI tutor's decisions at five decision points derived from the learning sciences (content complexity, cognitive task design, support structure, engagement mode, and disciplinary approach) against predetermined alignment matrices. In Part 1, we analyzed the model's behavior by auditing a frontier tutor across a factorial space of 2,000 learning scenarios, hereafter cases, applied PAI-corrective LoRA fine-tuning to the weakest cases on an open-weights proxy, and had trained educators review the original and fine-tuned outputs in a blind evaluation. In Part 2, we traced the corrections into the weights of the proxy, treating it as a model of the intervention rather than as a substitute for a commercial AI tutor, and we conducted a post hoc analysis to determine which conditions of a learning situation could be recovered from the cases by the models and from the text.

The study addresses two research questions.

RQ1: Does optimizing a metric of pedagogical adaptivity improve the instruction the metric measures?

- RQ1.1: How do trained educators' judgments correspond to the metric at baseline and after fine-tuning raises the metric?
- RQ1.2: Which property of the metric accounts for the observed relationship, and is that property separable from the metric gain?

RQ2: What changes inside the model when the metric is used for correction, and what determines where the changes take effect?

- RQ2.1: Where does the corrective update concentrate, and does its structure indicate a shallow, output-stage edit or a deeper representational change?
- RQ2.2: Do the metric-gain and sequence-level components of the correction occupy distinct directions in weight space?

Our contributions are geared toward the educational AI community and researchers in the field of responsible and explainable AI. This study shows that using a metric to improve an LLM's pedagogical behavior can degrade the quality the metric was designed to measure, and that automated evaluation alone cannot detect this degradation. Additionally, we identify a structural property of a metric that determines the safety of optimization, demonstrate the loss incurred when dynamic learning theories are decomposed into decision-point rules, and derive design principles for benchmarks that assess or train AI-tutor instruction.

## 2. Theoretical background

### 2.1 Pedagogical adaptivity as contingent instruction

The construct of pedagogical adaptivity rests on the foundational principle of the constructivist-cognitivist tradition that effective instruction is not a fixed set of practices, but rather a response that depends on the conditions under which learning occurs. This principle separates pedagogical expertise from content expertise. While a content expert knows the subject, a pedagogical expert knows how to teach it in different ways depending on who the learner is and where they stand in their learning trajectory.

The theoretical roots of adaptivity run deep, and each one grounds a decision point at which our index scores a tutor (Section 3.1). According to Vygotsky's (1978) Zone of Proximal Development, instruction is only effective when calibrated to the boundary between what a learner can do alone and what they can achieve with assistance, which Wood, Bruner, and Ross (1976) formalized as contingent scaffolding. Piaget's (1952) constructivism adds that the cognitive work of learning varies with the learner's existing schemas: constructing a new schema, assimilating compatible information, or accommodating knowledge that conflicts with a prior misconception (Posner et al., 1982; Chi, 2008). Chi and Wylie's (2014) ICAP framework categorizes interactive, constructive, active, and passive engagement by the depth of learning that each supports, and this ordering only applies when the engagement mode matches the task and the learner. Since no instructional policy is optimal across all learning conditions, pedagogical adaptivity can be defined as the degree to which an instructional agent calibrates its decisions to those conditions.

A point that is often overlooked when operationalizing adaptivity is that these conditions are not fixed properties of a learner. Rather, they are positions on a trajectory that instruction itself moves, as the source theories make explicit. Scaffolding is defined by contingency, fading, and the transfer of responsibility (van de Pol et al., 2010; Collins et al., 1989); support that never fades fosters dependence rather than scaffolding. Vygotsky's zone is a moving boundary. Piagetian accommodation names a transition, not a stable state. Even in the intelligent tutoring tradition, the knowledge state is modeled as evolving with each instructional act (Corbett & Anderson, 1994). Thus, competent instruction is contingent on a learner who changes precisely because instruction is happening, and a sequence of activities cannot be understood as a set of independent decisions calibrated to the same learner, but as a trajectory in which each activity is appropriate, in part, because of what preceded it.

For adaptivity to be measurable, an account of it must specify which conditions are important. We identified three learner conditions, each of which is a consensus dimension of the learning sciences, and each of which is understood as a position on a trajectory rather than a permanent label. The first condition is knowledge state, referring to the learner's knowledge relative to the target content (Ausubel, 1968); following schema theory, we classify learners as novice (with no relevant schema), informed (with a broadly accurate schema), or misinformed (with a conflicting schema that requires accommodation) (Posner et al., 1982; Chi, 2008). The second condition is the learning stage, which is the learner's position on the path toward competence, comprising conceptual orientation, skill building, competency development, and comprehensive mastery (Bloom et al., 1956; Dreyfus & Dreyfus, 1986; Belshaw, 2012). The third condition is the learning context, or the social setting that determines the tutor's role, which may be guided, collaborative, or autonomous (Vygotsky, 1978; Knowles, 1975; Moore, 1993).

Although specifying these conditions makes adaptivity measurable in principle, “appropriate” instruction admits no single correct answer. There are better and worse choices, as well as characteristics that appropriate instruction should exhibit under certain conditions. Our approach accordingly scores instruction at the level of characteristics rather than against a prescribed response, converting an unanswerable question into a tractable one. The operationalization of these principles and what it sets aside—the dynamics just described—is explained in Section 3.1.

### 2.2 Evaluating and optimizing LLM tutors

Recent work has considerably advanced the evaluation of LLM tutors, yet few studies target conditional adaptation as we have defined it here. Several benchmarks and rubrics assess pedagogical quality at the level of tutoring moves (Macina et al., 2025; Shi et al., 2026; Maurya et al., 2025). Another line of research consists of optimizing tutoring behavior through preference optimization and reinforcement learning using simulated student interactions, rewarding moves that an LLM judge scores as good scaffolding (Sonkar et al., 2024; Dinucu-Jianu et al., 2025). And our study examines the effects of such optimization when the score is a metric of adaptivity.

In these efforts, the focus of the evaluation is on a set of discrete, largely turn-level skills that are assessed independently of the learning conditions. Rubrics prioritize the tutor's side of the exchange over whether its decisions align with the learner's situation (LearnLM Team et al., 2025). What remains unexamined is whether a model conditions its instructional decisions on the learner's knowledge state, learning stage, and context, that is, whether it adapts not merely competently but appropriately to whom it is teaching and under what circumstances.

Automated scoring of adaptivity raises the question of how well such a metric corresponds to the judgments of trained educators, since pedagogical quality is ultimately anchored in expert judgment. Existing human–AI comparisons do not answer this question because they evaluate the two on different terms (Pal Chowdhury et al., 2025; LearnLM Team et al., 2025), and where automated and expert judgments have been compared directly, they diverge precisely where specialized expertise is required (Szymanski et al., 2025). In a systematic review of 445 benchmarks, Bean et al. (2026) found that construct validity was rarely established and proposed a checklist to secure it. This literature documents such divergences without characterizing their structure, that is, the pedagogical dimensions and conditions under which human and automated judgments diverge. Our first research question aims to establish that structure.

## 3. Methods

Both parts of the study are organized around a single instrument, which we present first and in full because its structure drives every result that follows.

### 3.1 The Pedagogical Adaptivity Index

The PAI was specifically designed for this study to measure the alignment between an AI tutor's response and a given learning situation. Derived from the theories reviewed in Section 2.1, it is built in the form used by the field for automated pedagogical scoring. It asks whether an instructional proposal exhibits the properties demanded by the situation's conditions, expressing the answer as a graded score on a scale from −1.0 to +1.0. We explain it in two phases: how instruction is elicited and how the AI tutor's response is scored.

#### *3.1.1 Elicitation*

Each learning situation is presented to the AI tutor as a prompt consisting of a case and a task (Figure 1). The case is a short narrative (80–120 words) describing the learner, their prior knowledge, their objective, and the study circumstances. The case implicitly includes five conditions: the three learner conditions from Section 2.1 (knowledge state, learning stage, and learning context); the cognitive level of the target objective, categorized from the revised Bloom taxonomy as remember/understand, apply/analyze, or evaluate/create (Anderson & Krathwohl, 2001); and the disciplinary family of the subject (formal, natural, humanistic, or applied). The task remains the same for each case: Design a sequence of five learning activities, and for each activity, make five selections and provide a brief description explaining each choice. These five selections correspond to five decision points, the dimensions along which the learning sciences hold that instruction must vary with conditions, and each point offers four ordered options spanning a recognized range of that dimension (Table 1). The options are ordered deliberately because appropriateness is graded rather than categorical. Responses are elicited under a constrained JSON schema so that selections can be parsed without inference. The index only scores the selections, while the descriptions were shown to the educators but do not contribute to any score.

Table 1. The five decision points of the PAI.

| **Decision point** | **Theoretical basis** | **Ordered options (a → d)** | **Scored against (matrices)** |
|---|---|---|---|
| DP1:<br>Content complexity | Vygotsky's ZPD (1978) | (a) Foundational exposition<br>(b) Structured explanation<br>(c) Analytical treatment<br>(d) Critical–advanced engagement | Cognitive level (A.1)<br>Knowledge state (A.2) |
| DP2:<br>Cognitive task | Piagetian schema operations (1952) | (a) Representation building<br>(b) Iterative practice<br>(c) Contextual bridging<br>(d) Connective synthesis | Learning stage (A.3)<br>Cognitive level (A.4) |
| DP3:<br>Support structure | Wood, Bruner & Ross scaffolding (1976) | (a) Directive guidance<br>(b) Complementary support<br>(c) Facilitative coaching<br>(d) Autonomy-supportive | Learning context (A.5)<br>Knowledge state (A.6) |
| DP4:<br>Engagement mode | Chi & Wylie's ICAP (2014) | (a) Receptive<br>(b) Manipulative<br>(c) Generative<br>(d) Collaborative | Learning context (A.7)<br>Cognitive level (A.8)<br>Learning stage (A.9) |
| DP5:<br>Disciplinary approach | Disciplinary epistemology (Kuhn, 1962; Becher & Trowler, 2001) | (a) Procedural–algorithmic<br>(b) Empirical–investigative<br>(c) Interpretive–argumentative<br>(d) Design–creative | Subject family (A.10)<br>Learning stage (A.11) |

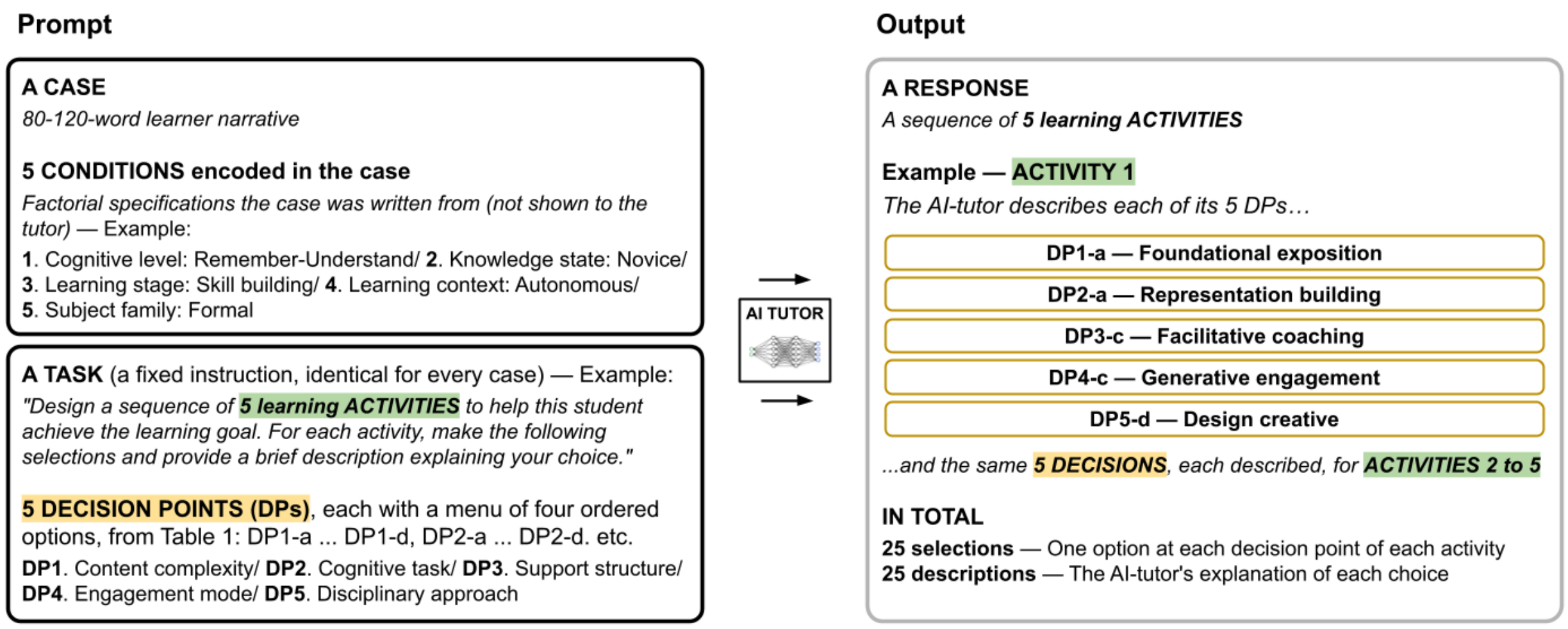

Figure 1. The elicitation process. The prompt pairs a case with a fixed task instruction that remains identical for every case, and the same prompt is used for both the audited model and the fine-tuned proxy. The figure shows Activity 1 of an actual response, including the option selected at each decision point and the tutor's description of it, with Activities 2 to 5 following the same form. Therefore, a response comprises 25 selections and 25 descriptions.

### *3.1.2 Scoring*

Scoring is defined at the level of a single decision and aggregated upward (Figure 2). Since not every decision depends on every condition, the framework pairs each decision point with the conditions identified by the learning sciences as governing it, yielding eleven alignment sub-matrices (Table 1, final column; Appendix A). Each matrix maps one option of one decision point under one value of one conditioning variable to a graded appropriateness score between −1.0 and +1.0. A decision score is the mean of the two or three matrix lookups that apply to it. From there, the index aggregates through the four levels shown in Table 2. Since every level is an unweighted mean of equal-sized groups, the scale of the matrices is maintained at each step and the aggregation is associative. Thus, a response's PAI is equal both to the mean of its 25 decision scores and to the mean of its five decision-point scores, making the decomposition by decision point exact.

Table 2. The levels of the index.

| Level | Name | Definition |
|---|---|---|
| 1 | Decision score | One selected option, at one decision point of one activity, scored against the conditions that govern that decision point — the mean of its two or three matrix lookups. |
| 2 | Activity score | The mean of that activity's five decision scores |
| 3 | PAI (of a response) | The mean of the response's five activity scores; the value reported for a tutor's answer to one case. |
| 4 | Mean PAI | The mean of response PAIs over a set of cases — the corpus, the intervention set, or the out-of-sample set. |
| — | Decision-point score | Cross-cutting: the mean of one decision point's scores across the activities of a response, and across cases when reported at corpus level. |

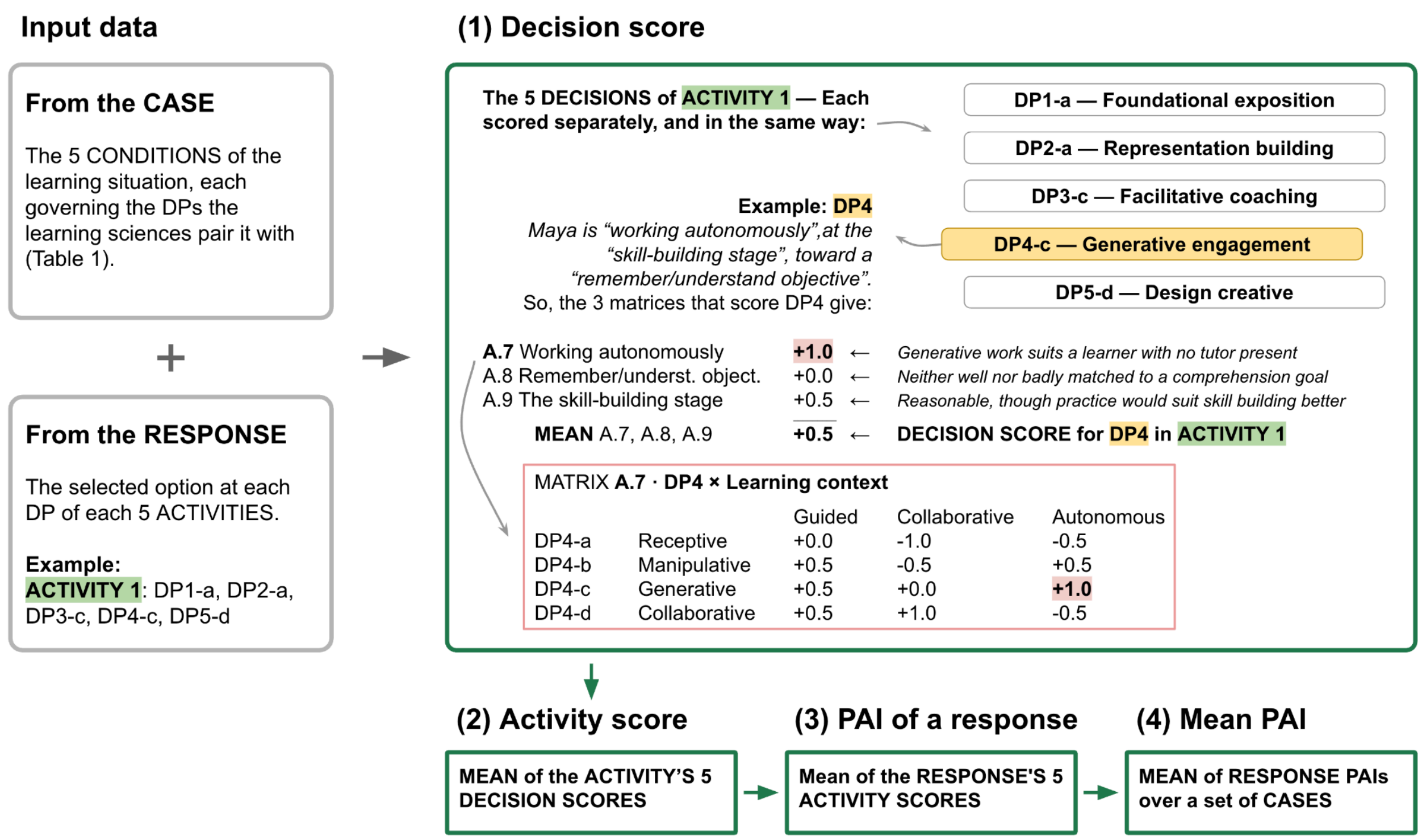

Figure 2. The scoring process. The inputs are the five conditions recovered from the case and the option selected at each decision point of each activity. The tutor's descriptions are not considered. Level 1 scores the five decisions of Activity 1 separately and expands one of them. For the engagement mode, the tutor chose generative engagement, and the decision score is the mean of the three matrices related to that choice.

### *3.1.3 Structural properties of the index*

There are two properties that are key to understanding the results and that are shared by every rubric-based pedagogical score we are aware of. First, each decision within an activity is scored independently, so the index contains no term relating choices across the five activities that comprise a sequence, thus providing no indication of whether a sequence varies, progresses, or coheres. Second, the conditions against which activities are scored are properties of the case and remain constant throughout the sequence. Therefore, the PAI is a separable objective, consisting of a mean of per-activity terms evaluated against a fixed context. This is the simplification that makes automated scoring possible, but it also means that the index can determine whether each decision fits the learner but not whether the sequence of decisions forms the kind of learning trajectory described in Section 2.1. Part 1 of this study examines whether that omission matters once the metric becomes a training target.

To make this limitation observable, we computed two sequence-level measures that are not part of the index and do not play a role in the corrective signal. Since the AI tutor selects one option at each decision point of every activity, each decision point yields five selections per response, one per activity. The repetition rate is the proportion of selections that are the modal option, and the selection entropy is the Shannon entropy of their distribution. Both are averaged over the five decision points to give one value per response. A response that repeats one option throughout has a repetition rate of 1.0 and an entropy of 0. These measures register whether a sequence varies at all, which is a property that the index cannot represent.

### 3.2 Research design

Part 1 comprises four experiments organized as a single design. The independent variable is corrective fine-tuning, with four levels: the original outputs of the audited model; the greedy PAI-optimal fine-tune, which is trained on targets that maximize the metric at every decision point; the diversity-preserving fine-tune, which is trained on targets that correct one decision point per example; and the base proxy, which is the un-fine-tuned proxy queried on the same cases. The dependent variables are the PAI, the two sequence-level measures, and the expert rating. Case type, distinguishing the 30 lowest-PAI cases targeted by the correction from 8 unmodified high-PAI cases, is a blocking factor. Experiment 1 is the baseline audit, Experiment 2 is the corrective fine-tuning, and Experiments 3 and 4 are the two rounds of expert evaluation. Part 2 comprises Experiments 5 and 6, which are two pre-committed experiments on the weights of the proxy.

Because the original outputs come from the audited model and the corrected outputs come from the proxy, the generator's identity could be confounded with the intervention. The base-proxy arm eliminates this confound by holding the generator fixed, maintaining identical cases, prompt, and decoding, while varying only the treatment. Every causal claim is made against this arm. We control for rater drift by leaving the control cases unchanged across both evaluation rounds, for order and demand effects by blinding participants and re-randomizing case order, and for adapter-capacity artifacts by sweeping the adapter rank. However, one threat could not be addressed: no educator rated the base proxy (Section 5.4).

### 3.3 Models and case corpus

The audited model is Gemini 3.1 Pro Preview (closed weights), which was chosen because the Gemini family incorporates LearnLM, Google's learning-tuned model line that is used in educational products (LearnLM Team et al., 2025). The proxy is Qwen3-32B (open weights), on which all fine-tuning is performed, and which is the focus of Part 2. Therefore, our mechanistic claims are about the effects of PAI-corrective fine-tuning on an open-weights, instruction-tuned transformer, not about any commercial AI tutor.

A full factorial crossing of the five conditions was sampled into 2,000 balanced scenarios, each of which was rendered as a case narrative and validated for faithfulness to its condition values, absence of answer-cueing hints, and de-duplication (Appendix B). The audited model received the fixed prompt for each scenario and provided a response as described in Section 3.1.1, which was then scored to yield a PAI, five decision-point scores, and the two sequence-level measures. Four case sets were drawn from this corpus. The intervention set comprises the 30 cases with the lowest PAI ($\leq 0.07$), and the control set comprises the 8 cases with the highest. Together, these 38 cases were the material for the expert evaluation. The corrective training set comprises 589 cases, which were stratified across the ten condition cells in which the audited model was weakest, with the intervention set included and the control set excluded. The out-of-sample set comprises 200 cases, which are also stratified across conditions and disjoint from the corrective set.

### 3.4 Corrective fine-tuning

Two forms of corrective supervision were created using the 589 training cases (Appendix B). The greedy signal selected the option that maximized the alignment-matrix scores under the case's conditions at every decision point of every activity, without considering any other criterion. Each selection was paired with a description drawn from a fixed bank of 100 learner-independent phrases (five paraphrases per option), so the five activities of a target had differently worded descriptions of the same choice. The diversity-preserving signal corrected a single decision point per example toward the metric, while leaving the remaining decisions as originally produced and retaining cross-activity variation. Four such adapters were built, one per corrected decision point. Both arms used LoRA on the proxy, fixing all hyperparameters except the rank of the greedy adapter, which was swept over $r \in \{1, 4, 8, 16\}$ under the standard convention of a fixed LoRA $\alpha$, resulting in an effective update scale of $\alpha/r$ that varies with rank. Each fine-tuned model, the base proxy, and the audited model were queried on the same cases and scored identically. The configuration, seeds, and released adapters are documented in Appendix B.

### 3.5 Expert evaluation

Thirty-one educators, out of the 51 who were invited, completed both rounds of the evaluation. All of them held a prior degree in education and were enrolled in a graduate AI-and-Education program. Participation was voluntary, ungraded, GDPR-compliant, and approved by the IRB. The 38 cases were rated in two rounds several weeks apart, under a balanced incomplete block design that assigned eight cases to each participant and allocated between ten and thirteen raters to each case. Given the completion rate, each case received between four and eleven independent ratings, with a mean of 6.5.

Each case was presented as the learner narrative, followed by the five activities proposed by the AI tutor, with the option selected at every decision point shown together with the tutor's description. Raters received a neutral, one-paragraph description of the five decision categories and rated six items on a scale from 1 to 5 (overall alignment and one item per decision point) before answering two free-text prompts identifying the weakest and strongest aspect of the plan (Appendix C). Experiment 3 rated the original outputs. Experiment 4 rated the same cases, substituting the greedy fine-tuned outputs ($r = 8$) for the intervention cases while leaving the control cases unchanged. The case order was randomized again between rounds, and the participants were blind to the index, to which cases had changed, and to the manipulation itself.

### 3.6 Representational analyses

Experiment 5 localizes the corrective update. For every adapted matrix, the update is $\Delta W = BA$. We computed its Frobenius norm, normalized it per element so that matrices of different sizes could be compared, and distributed it across network depth and module type (attention q/k/v/o, MLP gate/up/down, and output projection), together with the spectral profile and effective rank at every rank. The pre-committed interpretation is as follows: an update concentrated in the output stage that is achievable at a low rank and is nearly uniform across depth indicates a shallow behavioral edit, that is, a change in the outputs of the model rather than in the computation that produces them, whereas an update distributed across depth that engages directions not aligned with the dominant structure of the base model would indicate a deeper representational change.

Experiment 6 tests weight-space separability. For each adapted matrix, we computed the principal angles between the greedy and diversity-preserving update subspaces on the input (hidden-state) and output (vocabulary) sides and the fraction of the greedy update's energy within the diversity-preserving span.

Substantial overlap would mean that the metric gain and any accompanying degradation are carried by one direction and cannot be disentangled by better supervision, whereas substantial distinctness would mean that the two are separable in principle, so a corrective signal could target one direction without affecting the other.

### 3.7 Condition-recovery probe

Scoring a decision against a condition assumes that the model can recover that condition from the case. We tested this assumption in a subsequent analysis that was not part of the initial design (see Section 4.5 for motivation). We fed each of the 200 out-of-sample narratives back to the model and instructed it to classify the five conditions of the learning situation, selecting one option per condition under a constrained JSON schema. The five dimensions were presented in a fixed order for every case: cognitive level, knowledge state, learning stage, learning context, and subject family; therefore, a position effect cannot account for failure in any one of them. The option descriptions were derived from the coding scheme used to generate the corpus, ensuring that every ground-truth label was reachable, and the classifications were compared with the factorial specification from which each narrative originated. We report accuracy, the majority-class baseline, and Cohen's κ, which corrects for the unequal number of levels across conditions. Both the audited model and the proxy were probed with greedy decoding, and every response was parsed.

Since a condition that the models fail to recover may be absent from the narratives rather than unread by the models, we also estimated the recoverability of each condition from the text itself. We did this by training a bag-of-words classifier (using TF-IDF features and logistic regression) to predict each condition from the narrative under five-fold cross-validation on the 200 out-of-sample cases, and reporting its κ alongside the models'.

### 3.8 Data analysis

To address RQ1.1, we correlated per-case mean ratings with the PAI (Spearman) at each decision point and overall, tested whether educators could discriminate between intervention and control cases at baseline (Mann–Whitney U). After fine-tuning, per-case changes in PAI and rating were tested against zero (Wilcoxon), with controls serving as the no-change reference, and related to each other (Spearman). To address RQ1.2, we compared the PAI with sequence-level measures across the four arms. RQ2.1 and RQ2.2 are addressed in Experiments 5 and 6, respectively. Inter-rater agreement is reported per round as Krippendorff's α (ordinal), as well as the percentage of ratings that are 4 or 5, since α decreases under ceiling effects. The 992 free-text justifications (496 responses × two prompts) were coded using a deductive scheme with eleven codes: five for the decision points of the framework and six for properties outside the framework (Appendix D). A rule-based procedure was used to assign codes by matching each text against anchor phrases for every code. To validate the procedure, two educators independently coded a stratified 20% subsample of the weakest-aspect texts ($n = 99$), blind to the automated codes, and we report Cohen's κ between the two coders and between each coder and the procedure. Claims are restricted to codes on which both agreements are at least moderate.

# 4. Results

## 4.1 Baseline audit

Across 2,000 factorial scenarios, the audited model achieved a mean PAI of +0.29 (SD 0.11), with 96% of cases between 0 and +0.5 and nearly none negative (Figure 3a). The AI tutor rarely makes actively counterproductive decisions, nor does it often make optimal ones. When decomposed by learning condition, the mean PAI was lowest for novice learners at the skill-building stage, and the 30 cases with the lowest PAI that became the intervention set were dominated by that combination (Figure 3c).

Decomposing by decision point requires a correction because the alignment matrices differ in the number of rewarded options. The expected score of a uniformly random option is +0.246 at DP5, but only +0.135 at DP2. When measured against the random baseline (Table 3 and Figure 3b), adaptivity was weakest for support structure (DP3, +0.04 above random) and cognitive task design (DP2, +0.06), while the apparent strength of the disciplinary approach (DP5) decreased from +0.37 raw to +0.13 above random. Before fine-tuning, the proxy produces almost the same profile (ρ = 0.900; Table 3), so the profile is not specific to one model family.

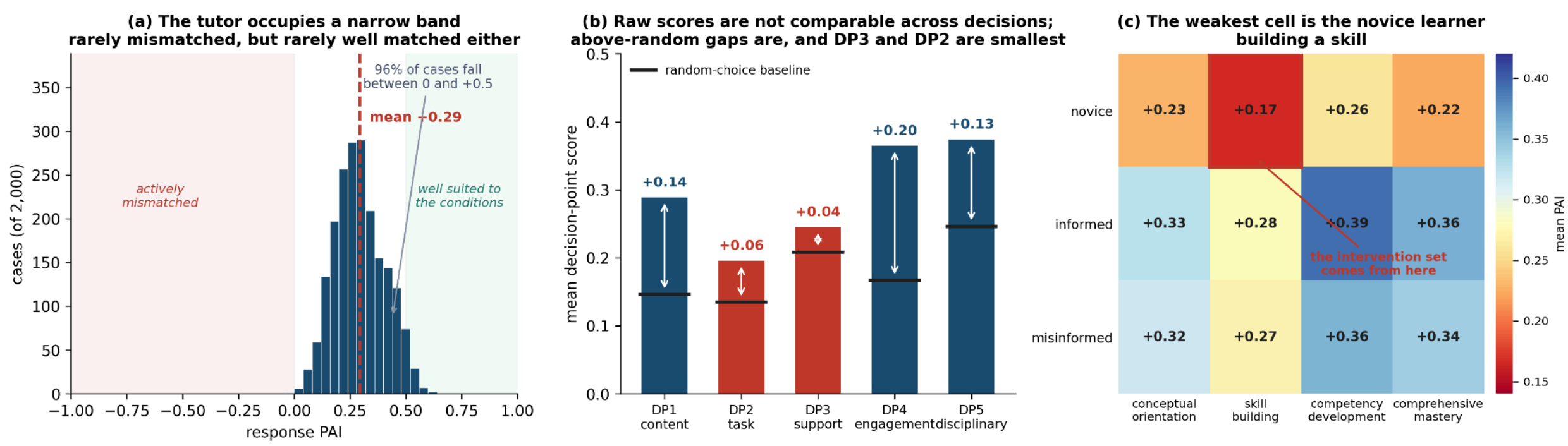

Figure 3. Baseline audit. (a) Distribution of response PAIs across the 2,000 scenarios on a −1 to +1 scale. (b) Mean decision-point score by decision point; the black rule inside each bar marks the expected score of a uniformly random option, and the arrow marks the gap above it. (c) Mean PAI by knowledge state and learning stage; the boxed cell is the source of the intervention set.

Table 3. Mean decision-point score in the baseline audit, before any fine-tuning, for both model families.

| Decision point | Gemini raw | Gemini above random | Qwen raw | Qwen above random | Random baseline |
|---|---|---|---|---|---|
| DP1 Content complexity | +0.289 | +0.143 | +0.116 | −0.030 | +0.146 |
| DP2 Cognitive task | +0.196 | +0.060 | +0.090 | −0.045 | +0.135 |
| DP3 Support structure | +0.245 | +0.036 | +0.149 | −0.060 | +0.209 |
| DP4 Engagement mode | +0.365 | +0.197 | +0.246 | +0.079 | +0.167 |
| DP5 Disciplinary approach | +0.374 | +0.128 | +0.325 | +0.079 | +0.246 |

Note. Audited model: 2,000 cases. Proxy: 589 cases. The random baseline for a decision point is the mean, across the 2,000 cases, of the average score of its four options under each case's conditions; only the above-random columns are comparable across decision points. Rank correlation between the two above-random orderings: Spearman ρ = 0.900.

### 4.2 Expert judgment against the metric, before and after correction

Regarding the original outputs (RQ1.1), the mean expert ratings per case were not related to the PAI (Spearman $\rho = -0.15$, $p = .39$), and no decision point reached significance ($|\rho| \leq 0.23$, all $p \geq .17$). Educators did not discriminate between the two ends of the metric either. The 30 cases with the lowest PAI scores and the 8 cases with the highest scores (mean PAI +0.05 and +0.59, respectively) received statistically indistinguishable ratings (4.46 vs. 4.39 on a scale from 1 to 5, $p = .53$). There was no inter-rater agreement (Krippendorff's $\alpha \approx 0$), reflecting a ceiling effect in which 90% of the ratings were a 4 or a 5.

After fine-tuning, the correction was effective on its own terms. The mean PAI rose from +0.05 to +0.42 on the 30 intervention cases, a gain of +0.37 that improved every decision point. Selected as the lowest-scoring in the corpus, these cases had the most room to improve, and the human evaluation judges the intervention under the conditions most favorable to it. Nevertheless, expert judgment moved in the opposite direction. In the same 30 cases, the mean overall rating decreased from 4.46 to 3.03 (Wilcoxon, $p < .00001$), and the decrease was significant for all six items (each $p < .00001$) (Figure 4b). The control cases, which were never modified, remained unchanged (4.39 to 4.51, $p = .38$), ruling out rater drift, fatigue, or a general shift in rating behavior. The metric and the educators were decoupled rather than inversely related, meaning that the size of the metric gain on a case did not predict how far its rating fell ($\rho = -0.24$, $p = .21$). As the ceiling effect dissipated, inter-rater agreement rose to $\alpha = 0.27$, with the share of ratings at 4 or 5 falling from 90% to 52%. Thus, educators agreed that the outputs were worse.

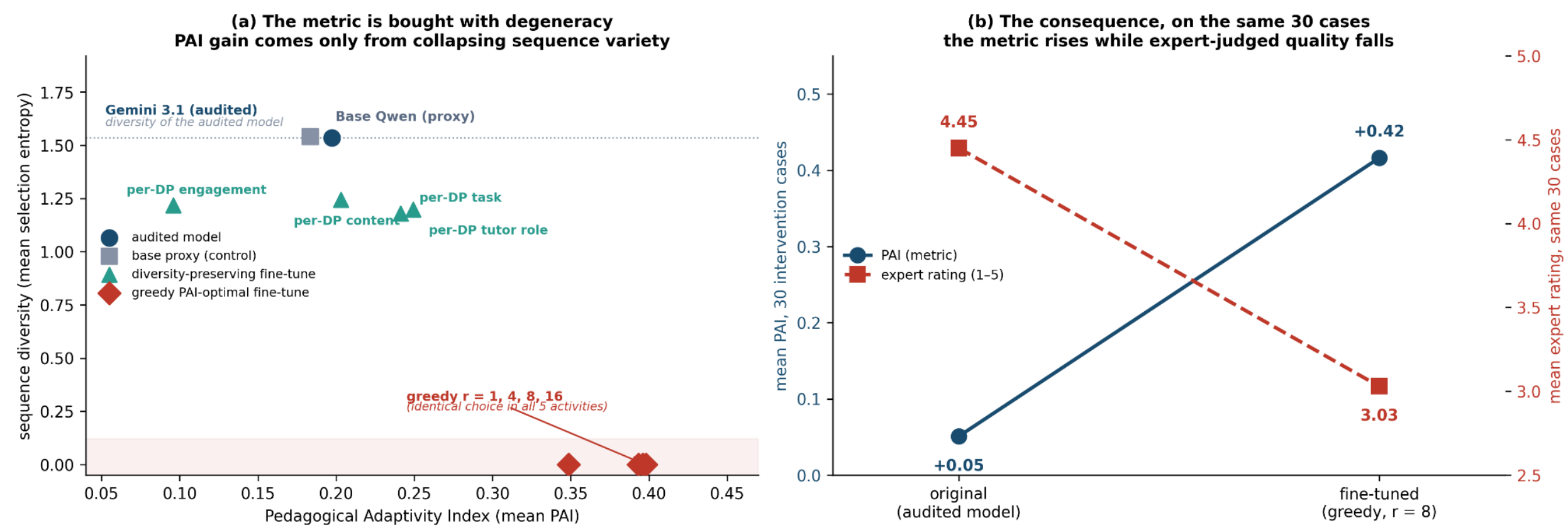

Figure 4. Trade-off. (a) PAI against sequence diversity (mean selection entropy) across all arms, on the 589 corrective cases: the audited model, the base proxy, the four diversity-preserving adapters, and the greedy adapters at $r$ = 1, 4, 8, and 16. (b) PAI and expert rating on the same 30 intervention cases, before and after greedy correction ($r$ = 8).

The analysis of the free-text justifications explains why the ratings decreased (Figure 5). Under human coding, approximately nine out of ten critiques of the weakest aspect addressed one of the five decision points, and the two coders substantially agreed on the tutor role and monotony codes ($\kappa = 0.65$ and 0.63) but less so on the remaining codes ($\kappa \leq 0.47$), so we only report what those two codes support. One property outside the framework was raised almost exclusively regarding the corrected outputs: monotony, or the repetition of the same option across the five activities. Both human coders identified monotony in 19–25% of weakest-aspect texts in corrected intervention cases and in at most 3% of texts on the original outputs and one control text. The automated procedure, which agreed with the coders at κ = 0.48–0.59 on this code, identified monotony in 16% of texts on corrected outputs and in none of the others. Monotony carried the largest penalty of any code: among corrected intervention outputs, responses whose justifications named it were rated 2.2, compared to 3.2 for those that did not (2.6 vs. 3.1 under automated coding). Educators penalized the very property that the metric had optimized into existence, even though the five descriptions accompanying each corrected activity were worded differently (Section 3.4); the monotony they identified was in the instructional choices, not the prose.

The corrected outputs differed from the originals in a second way that limits this result. Their descriptions were learner-independent phrases from the training bank that the fine-tuned model reproduced almost verbatim (Section 4.4). These descriptions were shorter than the originals (11.9 vs. 16.7 words) and no longer referred to the case: the share of description words also present in the learner narrative decreased from 0.21 to 0.03. Therefore, part of the rating decline may be attributable to generic descriptions rather than the collapsed selections, and the two cannot be fully separated in these data (Section 5.4). A change of generator alone does not produce either effect: the base proxy's outputs for the same cases had grounded descriptions (0.23) and no monotone sequences.

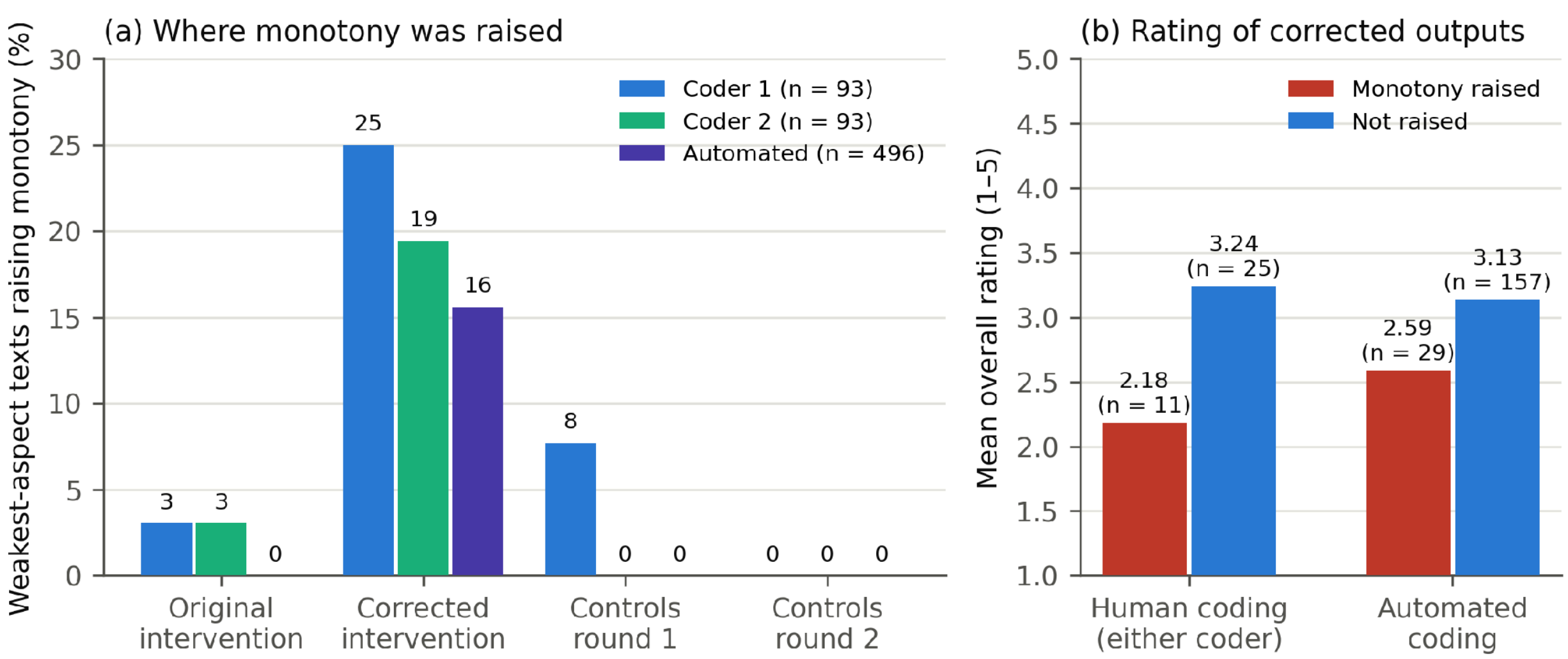

Figure 5. Monotony in the free-text justifications. (a) Proportion of weakest-aspect texts raising monotony, by round and case type, under each human coder and the automated procedure. (b) Mean overall rating of corrected intervention outputs with and without a monotony critique.

### 4.3 The degenerate optimum

To address RQ1.2, we examined the supervision that a metric of this form produces. Since each activity is scored independently against fixed conditions (Section 3.1.3), there is one optimal option for each decision point in each case, and the score-maximizing sequence of five activities is that optimal option repeated five times. Consequently, every corrective target is fully monotone (repetition rate 1.000, selection entropy 0.000), not due to how the targets were written, but due to the structure of the metric. The greedily fine-tuned models learned just that and produced fully monotone sequences in 100% of the cases tested at all four ranks (r = 1, 4, 8, 16). That the collapse is complete at rank one, where the update is carried by a single direction in each adapted matrix, shows that no high-dimensional change is needed to produce it.

The base-proxy control shows that the collapse is caused by the intervention and not by the change of model. On the 589 training cases, the un-fine-tuned proxy closely matched the audited model on the metric (mean PAI 0.184 vs. 0.197) and sequence diversity (repetition 0.490 vs. 0.486; entropy 1.540 vs. 1.536) with no monotone outputs. The same equivalence holds on the 30 intervention cases (PAI 0.059 vs. 0.051; repetition 0.481 vs. 0.483), on which the human comparison rests. The control cases, which were never corrected, are not part of this comparison. First, the fall in expert ratings cannot be attributed to generator quality since the two generators are indistinguishable in the very cases that the educators rated before treatment. Second, the low starting point of the intervention cases reflects the cases rather than an unlucky draw: a second model generating from scratch scores them at the same level, so the +0.37 gain reported in Section 4.2 is the effect of the correction, not regression to the mean.

The collapse generalizes out of sample. On 200 cases that were never seen during training, the greedily fine-tuned model again produced fully monotone sequences on 100% of the cases, with a mean PAI of 0.42, compared to the base proxy's mean PAI of 0.30. This shows that the degeneracy is a property of the intervention rather than the training set.

The described collapse occurs within a sequence, with the same option repeated across five activities. On cognitive task design (DP2), the collapse is also across cases. Each of the 200 unseen learners received the same option, DP2-b (iterative practice), regardless of knowledge state, stage, or context, whereas the base proxy applied four distinct options, and the fine-tuned model continued to vary its choices on the other four decisions (Figure 6b). In that dimension, the model no longer teaches differently to different learners. The metric itself registers the consequence. Decomposed by decision point, fine-tuning raised the out-of-sample decision-point score on four decisions, but lowered it on DP2 from 0.24 to 0.04—essentially the score of an uninformed choice—while the mean PAI nonetheless rose because gains on the other four decisions masked the loss (Figure 6a). Thus, a metric designed to measure conditional adaptation recorded an overall improvement while the model was eliminating conditional adaptation on one of the dimensions that the metric scores.

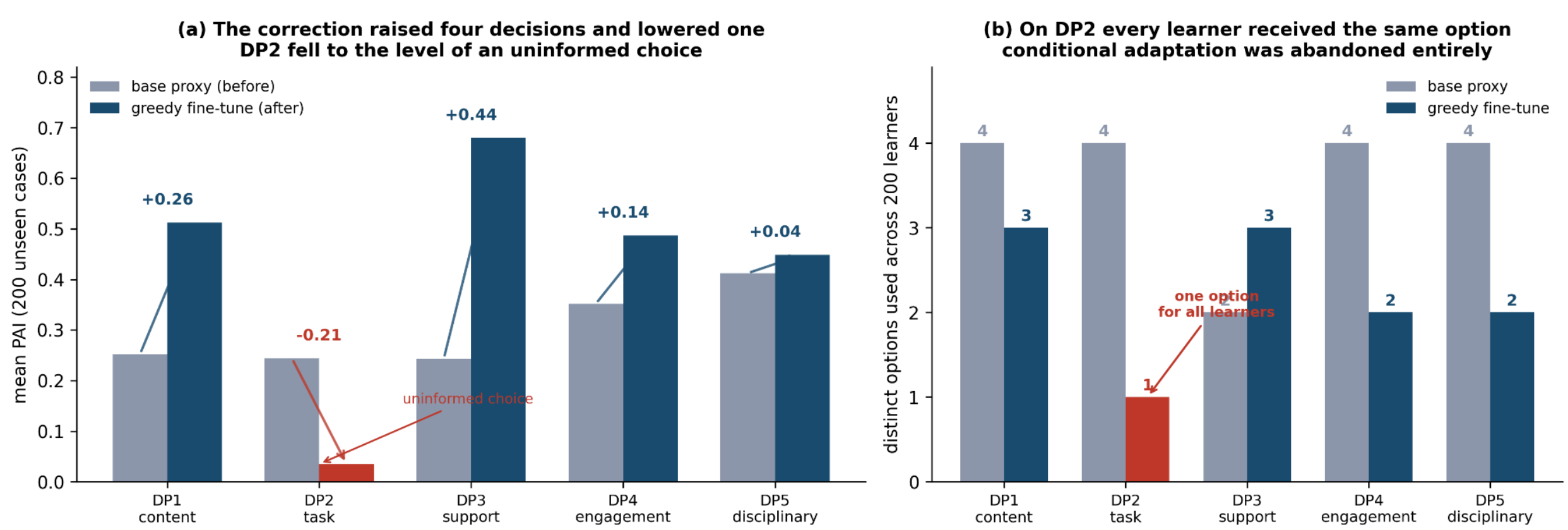

Figure 6. The effect of the correction, decomposed by decision point, on the 200 unseen cases. (a) Mean decision-point score before and after greedy fine-tuning. (b) Number of distinct options each arm applied across the 200 learners, taking the modal option per case as the policy that learner received.

The diversity-preserving arm shows what the metric offers when collapse is avoided. On the 589 corrective cases (Figure 4a), the base proxy scores 0.185 and the greedy model scores 0.40, while the diversity-preserving models only reach +0.20 (content), +0.24 (task), and +0.25 (support structure), and for engagement mode, the PAI falls to +0.10. The alternative recovers at most about a third of the gain. The metric's potential lies almost entirely in its degenerate region, meaning that, in practice, a high PAI can be achieved only by collapsing the sequence.

Finally, we tested whether restoring the dynamics the metric omits would serve as a remedy, but it does not. In a pre-specified rescoring, we allowed the learner's conditions to evolve across the sequence (the learning stage advancing one level every two activities, with the knowledge state moving to informed at the third) and scored each activity against the evolved conditions using the same matrices. Under this schedule, the advantage of the corrected outputs over the originals shrank from +0.37 to +0.07 (0.34 vs. 0.27, $p = .047$), and across a faster and a slower schedule it ranged from −0.02 ($p = .46$) to +0.15. It never reversed significantly, and the change in the rescored metric did not track the change in expert rating any better ($\rho = -0.09$). The rescored objective's optimum is no longer repetition (fully monotone in only 45 of the 432 condition cells, with no shared template); a third of its decisions are ties. Evolving conditions remove the degenerate optimum without enabling the metric to see what the educators saw.

### 4.4 Locating the correction in the weights

Weight-difference analysis (RQ2.1) unambiguously localizes the intervention. Between 75% and 93% of the update's energy is allocated to the output projection, which is the final map from hidden states to vocabulary logits ($r = 1$: 75%; $r = 4$: 92%; $r = 8$: 89%; $r = 16$: 93%). This is not a size artifact: when normalized per element, the output projection receives an update that is 18–30 times larger than any transformer block, while the 64 blocks are nearly uniform and almost flat across depth (23–31% of block energy per quartile compared to a 25% uniform baseline). Fine-tuning did not reshape the model's computation; rather, it rewrote the readout. The outputs show the same at the token level: for the intervention cases, the corrected model produced 748 verbatim strings from the 100-phrase training bank out of 750 descriptions. The fine-tuning led the model to memorize surface text rather than learn to describe instruction for a learner.

Consistent with this, the diversity-preserving adapters push measurably more of the change into the network's computation. Their output-projection share is lower (66–80%, and 41% for engagement mode), with greater involvement of the MLP blocks. Supervision that cannot be satisfied by a readout edit forces the model to change more of itself.

If the metric gain and collapse are separable, as the behavioral results suggest, they should be distinguishable in the weights (RQ2.2), and they are. Only 4.5% of the energy from the greedy update lies within the subspace spanned by the diversity-preserving updates overall. In the output projection, which carries most of the update, the two supervisions share roughly half of their output-side directions (48.8%) since they both push the same option tokens, which constitute the metric-gain component they have in common, but their input-side directions are nearly disjoint (only 3.3% are shared). The greedy update is triggered by features that are almost entirely different from those of the diversity-preserving update (Figure 7).

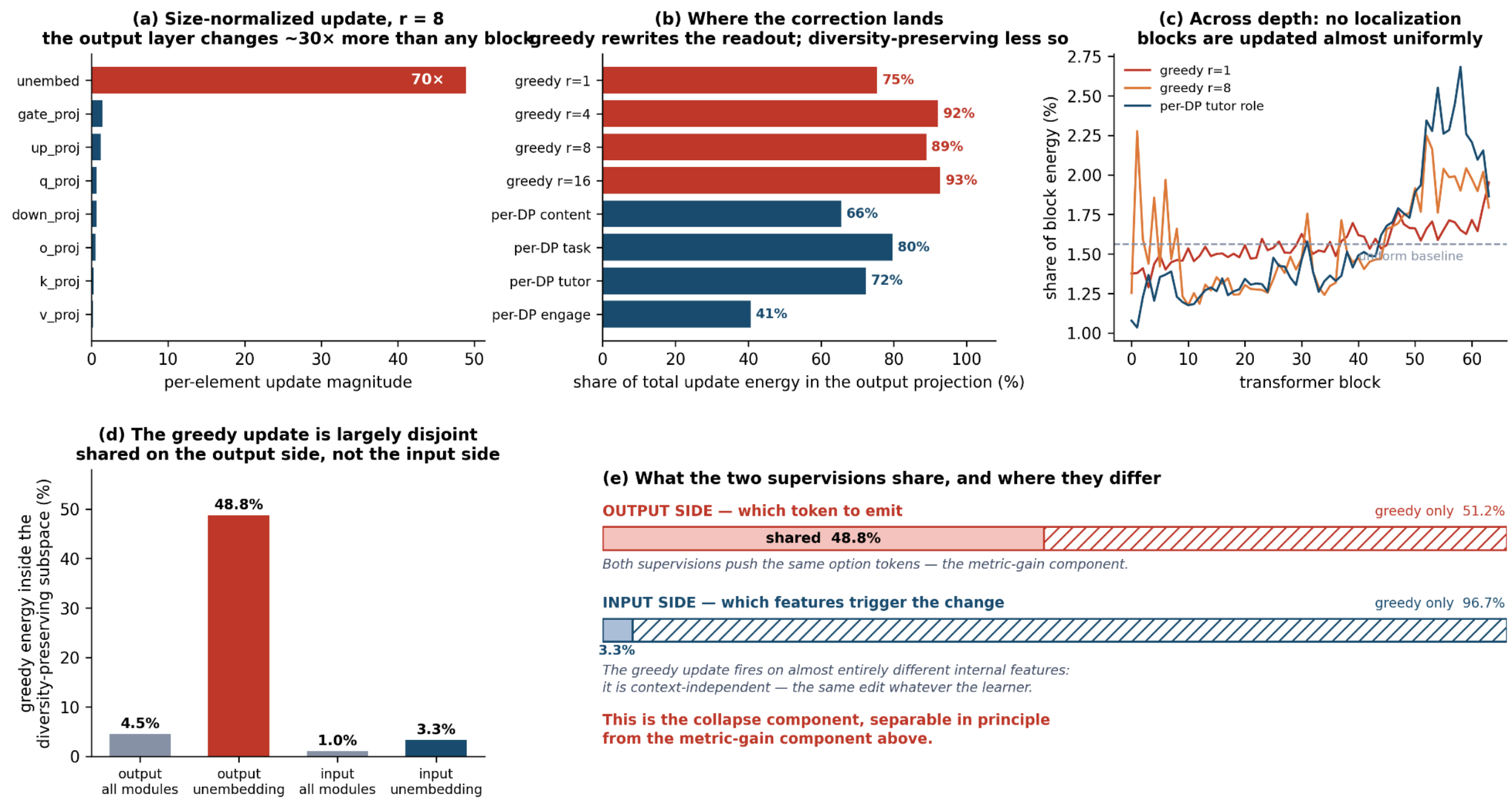

Figure 7. Where the correction landed in the weights. (a) Per-element update magnitude by module type at r = 8. (b) Share of total update energy in the output projection for the four greedy ranks and the four diversity-preserving adapters. (c) Share of block energy across the 64 transformer blocks against a uniform baseline. (d) Fraction of the greedy update's energy inside the diversity-preserving subspace, on the output and input sides. (e) Interpretation: the two supervisions share about half of their output-side directions, but almost none of their input-side ones.

### 4.5 Which conditions are recoverable

The correction did not affect all decision points equally. However, the metric's structure does not allow us to predict which decision point would give way, since separability applies identically to all five. One possible explanation lies in the source of the information the model uses to apply a condition at each point.

When we tested this explanation, we found that recoverability is not uniform (Table 4). Three of the five conditions are almost perfectly recovered by both models: subject family (κ = 1.00 for both, against the corpus's own taxonomy; Table 4), learning context (0.97 and 0.99), and knowledge state (0.98 and 0.93). The objective's cognitive level is recovered less well (0.73 and 0.66), and the learner's stage on the trajectory toward competence is not recovered at all (κ = 0.16 for the audited model and 0.05 for the proxy, whose 95% interval [−0.03, +0.11] contains zero). The failure to recover the learning stage is a systematic default, with predictions collapsing onto conceptual orientation regardless of the true stage, resulting in 38 of the 64 competency development cases and 28 of the 45 comprehensive mastery cases being labeled as conceptual orientation. The same asymmetry is present in the narratives themselves. A bag-of-words classifier can recover the learning context, cognitive level, subject family, and knowledge state from the text (κ = 0.98, 0.91, 0.86, and 0.74, respectively), but not the learning stage (κ = 0.15; Table 4). Thus, the models' default to the first stage reflects an uncertainty that the stimuli share, rather than an inability to interpret them, whereas the cognitive level, which the classifier reads better than either model, is a shortfall of the models themselves. In either case, the decisions scored against the learning stage were scored against a condition the model did not have.

Table 4. Recovery of each condition from the 200 out-of-sample narratives.

| **Condition** | **Majority baseline** | **Accuracy (audited)** | **Accuracy (proxy)** | **Accuracy (text classifier)** | **κ (audited)** | **κ (proxy)** | **κ (text classifier)** |
|---|---|---|---|---|---|---|---|
| Subject family | 0.405 | 1.000 | 1.000 | 0.910 | 1.000 | 1.000 | 0.862 |
| Learning context | 0.385 | 0.980 | 0.990 | 0.990 | 0.970 | 0.985 | 0.985 |
| Knowledge state | 0.410 | 0.985 | 0.955 | 0.835 | 0.977 | 0.930 | 0.737 |
| Objective cognitive level | 0.405 | 0.820 | 0.785 | 0.940 | 0.731 | 0.663 | 0.908 |
| Learning stage | 0.320 | 0.385 | 0.325 | 0.395 | 0.157 | 0.046 | 0.149 |

Note. κ is Cohen's unweighted kappa; the majority baseline is the frequency of the most common true label. The text classifier (TF-IDF features, logistic regression, five-fold cross-validation) estimates how recoverable each condition is from the narrative text itself. The subject family is scored against the corpus's own taxonomy (physics and psychology as natural sciences, sociology as applied).

When measured on the uncorrected models, recoverability predicts where the correction collapsed. Ranking the five decision points by the mean recoverability of the conditions that govern them yields the following order: DP3 (0.97), DP1 (0.83), DP4 (0.59), DP5 (0.55), and DP2 (0.40). Ranking them by the number of distinct options that the corrected model applied across 200 unseen learners yields the following order: 3, 3, 2, 2, and 1. These two orderings correlate at $\rho = +0.95$ ($p = .014$). However, the correlation with baseline adaptivity is much weaker at $\rho = -0.10$ ($p = .87$). Therefore, recoverability predicts vulnerability to metric-driven collapse, but not competence before the intervention. This is what a readout-level edit implies: a correction that fires regardless of context cannot use a condition that the model never recovered. DP2 is the only decision point with no governing condition recovered above $\kappa = 0.9$. Since the stimuli do not determine the learning stage, the baseline weakness of DP2 reported in Section 4.1 and the concentration of the intervention set in stage-dependent cells partly reflect the corpus rather than the models.

## 5. Discussion

The audit revealed that the tutor was weakly adaptive, and the correction increased the metric at every decision point while trained educators rated the corrected instruction lower in every category. We propose two ideas that explain the gap between metric-measured adaptivity and expert-judged alignment: the first concerns what is left out by these types of metrics, and the second concerns how little of the model is reached by a correction against them.

### 5.1 Measurement validity does not imply optimization validity

The PAI is derived from canonical educational theories and passes the checks that the field applies to a measure. Its matrices were frozen before any output was scored. It discriminated across the audit, and Bean et al.'s (2026) construct-validity checklist would have judged the metric favorably. The only relevant recommendation is a caution against single-point aggregation of labels, but this recommendation refers precisely to the aggregation that makes the objective separable. Thus, none of this makes a metric of this form safe to optimize, nor could anything in validating a measure do so. Construct validity is established against the distribution of existing behavior, whereas optimization drives behavior to the argmax, a region that no validation exercise visits. A checklist certifies a measure, not an objective.

This can be viewed as an educational version of Goodhart's law (Goodhart, 1975; Strathern, 1997), but with a more precise mechanism than the typical explanation, which states that a capable optimizer finds unexpected methods to satisfy an imperfect proxy (Gao et al., 2023; Karwowski et al., 2024). The failure reported here depends neither on the strength of the optimizer nor on a flaw in the content of the proxy. Rather, it follows from the algebraic form of the objective. A metric that independently scores each activity and then averages is maximized by repeating the single best-scoring activity. This is a pedagogically absurd sequence, yet it is the metric's true optimum. This degeneracy persists regardless of how well the individual judgments are calibrated, so no improvement to the alignment matrices could have prevented the collapse. Unhackability (Skalse et al., 2022) fails here for a reason visible in the objective's form rather than in the optimizer.

The absurd optimum has a pedagogical explanation: the operationalization retained the theories' contingency while discarding their dynamics (Section 2.1). However, by scoring each activity with conditions held constant across the sequence, the PAI encoded a learner who does not learn, for whom repeating the single best activity five times is correct by the metric's logic. Educators converged on a concern that the metric cannot represent: the plan repeats itself. The dynamics-preserving rescoring shows that this omission is harder to fix than it seems: allowing the learner's conditions to evolve eliminates repetition as the optimal solution, yet the rescored metric still favors the collapsed outputs over the originals and does not align with the educators' judgment. The educators penalized a property of the sequence as a whole, and a metric that reaches this property by adding a trajectory to a per-activity score has not yet been shown to exist.

### 5.2 The correction only affected the readout

The correction rewrote the readout rather than the computation. This is consistent with a broader pattern in fine-tuning wherein newly injected knowledge is stored in the weights yet is not routed into the circuits required for reasoning, so that the failure to generalize is a routing problem rather than a capacity issue (Dai et al., 2026). A context-independent readout edit must stop conditioning somewhere. DP2 was the decision point that gave way; it was also the weakest decision point at baseline. Conditioning was the most difficult here because it is scored against both the objective's cognitive level, which the models read less well than the text allows, and the learner's stage, which the narratives did not carry (Section 4.5). The audited model's difficulty with the analytical and evaluative levels of Bloom's taxonomy reproduces the pattern from our earlier audit (Domínguez et al., 2026) using a different metric, although the proxy does not show it, and is consistent with the finding that evaluation suites cluster in the lower tiers of the taxonomy (Huber & Niklaus, 2025), suggesting that the field may be assessing the models' weakest areas. Since both audits operationalize cognitive demand through Bloom's taxonomy, the shared method could account for part of the convergence.

The two supervision arms make this a testable claim. When the corrective targets could be satisfied by a context-independent edit, the update concentrated in the readout (75–93% of its energy). When they could not be satisfied, however, the update migrated measurably into the network's computation (66–80%, and 41% for the arm whose targets varied the most). Thus, the depth of representational change tracked the demands of the objective rather than the capacity of the model. The claim concerns what fine-tuning changed, not what the base model is. In fact, the un-fine-tuned models produced varied, condition-sensitive instruction. Our design does not address whether that adaptivity is based on representations of the learner's conditions or on shallower associations.

### 5.3 Implications for benchmarks of AI-tutor instruction

The PAI shares its form with the rubrics that the field uses to evaluate and increasingly to train LLM tutors. These rubrics score moves independently against fixed criteria and aggregate them by averaging (Maurya et al., 2025; LearnLM Team et al., 2025; Sonkar et al., 2024; Dinucu-Jianu et al., 2025). The findings demonstrate a property of that form on a metric better grounded than most, and each finding implies a design rule for benchmarks intended to assess or train instruction (Table 5). These rules expand upon the construct-validity checklist of Bean et al. (2026) by introducing a new concept: optimization validity. Four of these rules—verifying recoverability, separating recognition from decision, measuring adaptation as a difference, and publishing the argmax—are not, to our knowledge, mentioned in the current benchmark literature.

Table 5. From findings to design principles for benchmarks of AI-tutor instruction.

| Finding | Design principle |
|---|---|
| One conditioning variable was neither recoverable by the models nor from the stimulus text, so the collapse fell on the decision scored against it. | Verify that each conditioning variable can be recovered from the stimuli before scoring against it, and publish recoverability per condition as benchmark metadata. |
| Three conditions were read nearly perfectly, one was not read at all, and the decision failure occurred where reading failed. | Add a condition-recognition subtask to the instructional-decision task so that failure can be attributed to reading the learner or acting on the reading. |
| The most penalized property was monotony, a quality that no move-level rubric can see. | Score sequence-level properties (variation, progression, and fading) as dimensions in their own right. |
| On DP2, the corrected model provided the same option to every learner, and the aggregate score increased. | Measure adaptation as a difference by pairing items that differ in one learner condition and scoring the change in the tutor's decision. |
| The metric's optimum was one activity repeated five times. | Publish the argmax to show what a perfect score looks like. A repeated activity or a fixed template marks the benchmark as unsafe to optimize. |
| The collapse was registered by withheld measures, not the metric. | Report diagnostics that are excluded from the score, such as repetition, entropy, and grounding, on the same leaderboard. |
| The apparent strength of DP5 was halved against its random baseline. | Report above-random baselines per dimension. Raw scores are not comparable across rubric dimensions with different numbers of rewarded options. |
| $\alpha$ is approximately 0 at the ceiling and 0.27 once outputs are degraded. | Anchor expert judgment on outputs that span quality, including deliberately degraded ones. |
| Within-band recovery ranged from 0.59 to 0.97. Models treat Analyze and Evaluate differently from Create. | Do not band cognitive levels. |

### 5.4 Limitations and conclusions

Since the audited model's weights are closed, the representational analysis used an open-weights proxy. The base-proxy control behaves indistinguishably from the audited model on the metric and on sequence diversity, both on the 589 training cases and on the 30 intervention cases. However, no educators rated the base proxy, so the causal claim for human evaluation relies on computational equivalence rather than human ratings.

The corrected outputs the educators rated differed from the originals in two ways: collapsed selections and generic descriptions drawn from a fixed bank, and the design cannot fully apportion the rating decline between them. The monotony judgment did not depend on the prose (Section 4.2), but the size of the overall decline should be considered in light of the second difference. The confound would be removed by a replication in which corrected selections are paired with the model's own descriptions.

The human evaluation is in-sample; the 30 intervention cases are within the corrective training set. Thus, educators judged the model on cases on which it was trained (Section 4.2). The behavioral collapse, by contrast, generalizes out of sample.

The qualitative coding was rule-based and validated by only two coders, who agreed substantially on two codes; the raters were graduate students rather than in-service teachers, and the case narratives are synthetic. They were validated for faithfulness to their conditions by the generation pipeline, but one condition was weakly encoded, so synthetic stimuli should be validated by humans before they are scored. Pedagogical quality is realized over the trajectory of an actual learner whose knowledge state changes and whose misconceptions only surface under the right task, so an evaluation that abstracts this away risks certifying capabilities that do not survive real instruction.

Separability is established descriptively. The metric-gain and collapse components occupy distinct subspaces. However, we have not yet demonstrated that removing the orthogonal component preserves the gain without the collapse. The next step is to project the greedy update onto and off the shared subspace and measure the behavioral result. The diversity-preserving targets are not released in their training form; only the outputs and weights of the four adapters are released, so that arm's supervision cannot be independently verified.

We did not fix the training seed, and each condition was trained once, so we cannot distinguish differences between ranks from initialization noise, a caveat compounded by the fixed-α convention under which rank covaries with update scale. However, the principal finding does not rest on such differences: the collapse is total at every rank and is reproduced on 200 held-out cases. Additionally, the index groups Bloom's six levels into three bands, and the probe shows that the audited model recovers levels within a band at noticeably different rates; therefore, the results depend on the banding.

Two extensions would address these constraints: replicating the pipeline on a second open-weights model to determine whether metric-optimization collapse is a property of the objective, as our analysis suggests, and basing the case narratives on a real curriculum. In this case, in-service teachers would serve as evaluators, and sequence-level items would be built into the instrument from the beginning.

In sum, the metric improved, but the tutor performed worse. This study proved that any metric that scores instructional decisions independently is maximized by repeating the best decision, and optimizing it drove the model into that degeneracy on cases it was trained on and on cases it had never seen. Trained educators identified the repetition that the metric could not represent, and the weights showed that the correction rewrote the model's output rather than its reasoning. Measurement validity therefore does not imply optimization validity: a metric can accurately describe existing behavior and still be detrimental to optimize, because optimization leads to regions that are never inspected by validation. When operationalizing the learning sciences, the dynamics that make instruction contingent on a changing learner must be preserved, and our rescoring shows that adding them back to a per-activity score is insufficient. This work points toward benchmarks built according to the principles in Table 5 and toward evaluation embedded in real learning settings, where the alignment of an AI tutor with a learner who changes can be observed directly.

## Appendix A. The PAI alignment matrices

Each table gives the appropriateness score of the four options of one decision point under each value of one conditioning variable. A decision score is the mean of the two or three tables that apply to it (Table 1, final column).

Table A.1. DP1 Content complexity × Objective cognitive level (Bloom band).

| Option | RU | AA | EC |
|---|---|---|---|
| (a) | +1.0 | -0.5 | -1.0 |
| (b) | +0.5 | +0.5 | -0.5 |
| (c) | -0.5 | +1.0 | +0.5 |
| (d) | -1.0 | +0.5 | +1.0 |

RU = Remember/Understand; AA = Apply/Analyze; EC = Evaluate/Create.

Table A.2. DP1 Content complexity × Knowledge state.

| Option | novice | informed | misinformed |
|---|---|---|---|
| (a) | +1.0 | +0.0 | -0.5 |
| (b) | +0.5 | +1.0 | +0.0 |
| (c) | -0.5 | +0.5 | +0.5 |
| (d) | -1.0 | -0.5 | +1.0 |

Table A.3. DP2 Cognitive task × Learning stage.

| Option | conceptual orientation | skill building | competency development | comprehensive mastery |
|---|---|---|---|---|
| (a) | +1.0 | -0.5 | -0.5 | -0.5 |
| (b) | -0.5 | +1.0 | +0.0 | -0.5 |
| (c) | +0.0 | -0.5 | +1.0 | +0.5 |
| (d) | -0.5 | -0.5 | +0.5 | +1.0 |

Table A.4. DP2 Cognitive task × Objective cognitive level (Bloom band).

| Option | RU | AA | EC |
|---|---|---|---|
| (a) | +1.0 | +0.0 | -0.5 |
| (b) | +0.5 | +0.5 | -0.5 |
| (c) | -0.5 | +1.0 | +0.5 |
| (d) | -1.0 | +0.5 | +1.0 |

RU = Remember/Understand; AA = Apply/Analyze; EC = Evaluate/Create.

Table A.5. DP3 Support structure × Learning context.

| Option | guided | collaborative | autonomous |
|---|---|---|---|
| (a) | -0.5 | -1.0 | +1.0 |
| (b) | +1.0 | +0.0 | +0.0 |
| (c) | +0.5 | +1.0 | -0.5 |

| Option | guided | collaborative | autonomous |
|---|---|---|---|
| (d) | -0.5 | +0.5 | +0.5 |

Table A.6. DP3 Support structure × Knowledge state.

| Option | novice | informed | misinformed |
|---|---|---|---|
| (a) | +1.0 | +0.0 | +0.5 |
| (b) | +0.5 | +0.5 | +0.0 |
| (c) | -0.5 | +0.5 | +1.0 |
| (d) | -1.0 | +1.0 | -0.5 |

Table A.7. DP4 Engagement mode × Learning context.

| Option | guided | collaborative | autonomous |
|---|---|---|---|
| (a) | +0.0 | -1.0 | -0.5 |
| (b) | +0.5 | -0.5 | +0.5 |
| (c) | +0.5 | +0.0 | +1.0 |
| (d) | +0.5 | +1.0 | -0.5 |

Table A.8. DP4 Engagement mode × Objective cognitive level (Bloom band).

| Option | RU | AA | EC |
|---|---|---|---|
| (a) | +1.0 | -0.5 | -1.0 |
| (b) | +0.5 | +0.5 | +0.0 |
| (c) | +0.0 | +1.0 | +0.5 |
| (d) | -0.5 | +0.5 | +1.0 |

RU = Remember/Understand; AA = Apply/Analyze; EC = Evaluate/Create.

Table A.9. DP4 Engagement mode × Learning stage.

| Option | conceptual orientation | skill building | competency development | comprehensive mastery |
|---|---|---|---|---|
| (a) | +0.5 | -0.5 | -0.5 | -1.0 |
| (b) | -0.5 | +1.0 | +0.0 | -0.5 |
| (c) | +0.5 | +0.5 | +1.0 | +0.5 |
| (d) | +0.0 | -0.5 | +0.5 | +1.0 |

Table A.10. DP5 Disciplinary approach × Subject family.

| Option | formal | natural | humanistic | applied |
|---|---|---|---|---|
| (a) | +1.0 | -0.5 | -1.0 | +0.0 |
| (b) | +0.0 | +1.0 | +0.0 | +0.5 |
| (c) | -0.5 | +0.0 | +1.0 | +0.5 |
| (d) | +0.5 | +0.5 | +0.5 | +0.5 |

Table A.11. DP5 Disciplinary approach × Learning stage.

| Option | conceptual orientation | skill building | competency development | comprehensive mastery |
|---|---|---|---|---|
| (a) | -0.5 | +1.0 | +0.5 | +0.0 |
| (b) | +1.0 | +0.5 | +0.0 | +0.0 |
| (c) | +0.5 | -0.5 | +0.5 | +0.5 |
| (d) | +0.0 | -0.5 | +0.5 | +1.0 |

**Provenance.** The matrices were specified a priori from the source theories (Section 2) before any model output was scored, and were frozen for the duration of the study. They encode graded appropriateness rather than a single correct answer: under most conditions several options score positively.

# Appendix B. Models, data, and training configuration

### B.1 Models

| Role | Model | Used in |
|---|---|---|
| Case-narrative generator | grok-3-fast (xAI) | Corpus generation |
| Audited tutor | gemini-3.1-pro-preview (closed weights) | Experiments 1 and 3 |
| Proxy tutor | Qwen/Qwen3-32B (open weights) | Experiments 2, 4, 5, and 6 |

The proxy was fine-tuned with LoRA on Tinker and served for evaluation through vLLM.

### B.2 Case corpus

A full factorial crossing of five conditions — objective cognitive level (Bloom band: RU / AA / EC), subject family (formal / natural / humanistic / applied), knowledge state (novice / informed / misinformed), learning stage (conceptual orientation / skill building / competency development / comprehensive mastery), and learning context (guided / collaborative / autonomous) — was sampled to 2,000 balanced scenarios.

Each scenario was rendered as an 80–120-word case narrative by grok-3-fast (temperature 0.8, maximum 300 tokens), with the condition values embedded implicitly rather than stated. Narratives passed a validation and de-duplication pass checking faithfulness to the assigned condition values and the absence of lexical cues that would signal a pedagogically expected answer.

**Case sets.**

| Set | n | Definition |
|---|---|---|
| Corrective training set | 589 | Stratified across the ten condition cells in which the audited model was weakest (75 cases per cell; 52 for the smallest); a case may serve more than one cell. |
| Intervention set | 30 | Lowest-PAI cases (PAI ≤ 0.07); contained within the corrective set. |
| Control set | 8 | Highest-PAI cases; excluded from the corrective set. |
| Out-of-sample set | 200 | Stratified across conditions; disjoint from the corrective set. |

### B.3 Corrective supervision

**Greedy (metric-optimal).** For each training case, the option maximizing the PAI sub-matrix scores at every decision point of every activity, paired with a description drawn from a fixed bank of 100 learner-independent phrases (five paraphrases per option). Monotone by construction (repetition 1.000, selection entropy 0.000).

**Diversity-preserving (per decision point).** Four corrective sets, each correcting a single decision point toward the metric while leaving the remaining decisions as originally produced, retaining cross-activity variation (repetition ≈ 0.60, entropy ≈ 1.20).

### B.4 Fine-tuning configuration

LoRA training on Tinker, targeting all linear modules (attention q, k, v, o; MLP gate, up, down; and the unembedding), giving 449 adapted matrices per adapter.

| Parameter | Value |
|---|---|
| Base model | Qwen/Qwen3-32B |
| Optimizer | Adam ($\beta_1 = 0.9$, $\beta_2 = 0.95$, $\varepsilon = 1 \times 10^{-12}$, weight decay 0.0, no gradient clipping) |
| Learning rate | $1 \times 10^{-4}$, constant (no warmup, no decay) |
| Epochs | 3 |
| Batch size | 1 sequence (one optimizer step per example) |
| LoRA rank | $r \in \{1, 4, 8, 16\}$ for the greedy sweep; r = 8 for the decision-point adapters |
| LoRA α | 32, fixed across all ranks |
| LoRA dropout | 0.0 |
| rsLoRA | Disabled, so the effective scaling factor is α/r |
| Loss | Cross-entropy over the final assistant message, mean reduction |
| Random seed | Not fixed (see §B.7) |
| Total steps (r = 8) | 1,767 (589 examples × 3 epochs) |

**Training outcomes.**

| | r = 1 | r = 4 | r = 8 | r = 16 |
|---|---|---|---|---|
| Final training loss | 0.0028 | 0.0031 | 0.0029 | 0.0030 |
| Wall-clock | 2.56 h | 3.21 h | 2.75 h | 1.72 h |

Decision-point adapters (all r = 8, 3 epochs, lr $1 \times 10^{-4}$): content complexity, n = 150, final loss 0.0124; cognitive task, n = 300, loss 0.0120; support structure, n = 225, loss 0.0115; engagement mode, n = 52, loss 0.0168.

**A note on fixed α.** Because α is held at 32 while r varies, the effective LoRA scaling factor α/r differs across the sweep: 32 at r = 1, 8 at r = 4, 4 at r = 8, and 2 at r = 16. This is the standard Tinker convention (not an anomaly), but it means the rank sweep covaries adapter capacity with update scale. Claims resting on the sweep are accordingly framed as concerning the rank sweep as conventionally configured, not the isolated effect of capacity.

Adapter weights for all ranks and all decision-point variants are released as safetensors, together with their adapter_config.json files and training-job identifiers.

### B.5 Decoding parameters

Generation settings differ by experiment and are reported here per arm, because sequence diversity is a dependent variable and is sensitive to sampling.

All model queries used a constrained JSON schema, so option selections were parsed without inference.

| Arm | Model | Temperature | Max tokens | Other |
|---|---|---|---|---|
| Corpus generation | grok-3-fast | 0.8 | 300 | — |
| Experiment 1 — audited tutor (2,000 cases) | gemini-3.1-pro-preview | 0.7 | 8,192 | thinking_level: high; structured output via response_schema and response_mime_type |
| Experiment 2 — proxy arms on the 589 corrective cases (base proxy; greedy adapters at r = 1, 4, 8, 16; decision-point adapters) | Qwen3-32B ± LoRA, via Tinker | 0.7 | 4,096 | top_p = 1.0, top_k disabled, one sample per case, Qwen3 stop sequences |
| Experiments 2 and 4 — proxy arms on the 200 out-of-sample cases and the 38 rated cases (base proxy; greedy r = 8 on the 30 intervention cases) | Qwen3-32B ± LoRA, via vLLM | 0.0 (greedy) | 4,096 | Deterministic decoding |

**The collapse is not a decoding artifact.** The two proxy evaluation paths used different decoding regimes, and the degeneracy appears under both. On the 589 corrective cases, sampled at temperature 0.7, every greedily fine-tuned model produced fully monotone sequences on 100% of cases at every rank, while the base proxy retained full option diversity (mean selection entropy 1.540) under the identical settings. On the 200 out-of-sample cases under greedy decoding, the same collapse recurred at 100%. Repetition therefore cannot be attributed either to a deterministic decoder or to a particular sampling temperature: the corrected model repeated a single option across all five activities even where the sampler was free to vary them.

### B.6 Reproducibility

All reported analyses are computed from the released adapters, datasets, and scoring code rather than from re-running training. Evaluation decoding on the out-of-sample and rated-case arms is greedy and therefore deterministic; the absence of a decoding seed does not affect reproducibility of those results.

### B.7 Known limitation: training seed

The training seed was not fixed, so LoRA initialization was random and each condition was trained once. As a consequence, training is not bit-for-bit reproducible from the released configuration, though the released adapter weights reproduce every reported analysis exactly. With a single run per condition there is no estimate of run-to-run variance, so small differences between ranks cannot be distinguished formally from initialization noise. The principal finding does not rest on such differences: the collapse is total (100% of cases) at every rank including rank one, and it reproduces on 200 cases disjoint from the training set, which makes a fortunate single run an implausible explanation.

### B.8 Note on training loss

Final training loss converged to approximately 0.003 at every rank, indicating that the model had very nearly memorized the 589 corrective targets. Because those targets are monotone by construction, near-perfect fit to them and the observed sequence collapse are two descriptions of the same outcome. This is consistent with the localization result of Experiment 5, in which the update concentrates in the output projection: fitting a fixed set of target tokens is achievable by a readout edit and does not require the model to alter the computation that selects them.

## Appendix C. The expert-rating instrument

Educators rated each case on six items (1 = not at all appropriate, 5 = fully appropriate) and answered two free-text prompts. The six Likert items are worded in ordinary pedagogical language; the mapping to the PAI decision points, used only in analysis and never shown to raters, is given below.

| Item | Wording (rated 1–5) | Maps to |
|---|---|---|
| 1 | Overall alignment: Considering this learner's specific situation, needs, and goals, how well does the AI tutor's instructional plan align with what would be pedagogically appropriate? | Overall PAI |
| 2 | Difficulty and depth: Is the complexity and depth of the content well-suited to this learner's current level of knowledge and preparation? | DP1: content complexity |
| 3 | Learning tasks: Are the cognitive tasks the student is asked to perform well-suited to what this learner needs to accomplish at this stage of their learning? | DP2: cognitive task |
| 4 | Guidance and support: Is the level of guidance, support, and interaction the AI tutor provides well-suited to this learner's situation and degree of autonomy? | DP3: support structure |
| 5 | Student engagement: Is the way the student is expected to engage with the material and with others well-suited to their learning context and goals? | DP4: engagement mode |
| 6 | Subject-matter approach: Is the methodological approach to the discipline well-suited to the subject and the learning objectives? | DP5: disciplinary approach |

**Free-text prompts.**

- Weakest aspect: “Which aspect of this instructional plan is the weakest for this particular learner? Please explain why.”
- Strongest aspect: “Which aspect of this instructional plan is the strongest for this particular learner? Please explain why.”

**Presentation.** Each case showed the learner narrative followed by the tutor's five-activity plan, each activity displaying its selected option at each decision point together with the model's description. Raters received a neutral one-paragraph description of the five decision categories. They were not shown the PAI, any score, the alignment matrices, or the intervention/control status of a case. Cases were delivered by Google Forms; each participant received 8 cases under a balanced incomplete block design that allocated ten to thirteen raters per case; with 31 of the 51 invited educators completing both rounds, each case received between four and eleven independent ratings (mean 6.5).

## Appendix D. Qualitative codebook

The free-text justifications were analyzed with a codebook developed through a hybrid deductive–inductive content analysis (Braun & Clarke, 2006; Hsieh & Shannon, 2005). The deductive layer (Family A) maps educator attention onto the five PAI decision points, measuring how much of what educators say falls within the framework the metric operationalizes. The inductive layer (Family B) captures properties educators raise that the PAI does not measure — the candidate explanation for the metric–human dissociation. Each response is one unit of analysis and may receive multiple codes; codes were assigned by a rule-based procedure that matches each text, in Spanish or English, against anchor phrases for every code.

- **Family A — within-framework attention.** A1 content level / difficulty (↔ DP1); A2 cognitive task type (↔ DP2); A3 tutor role, guidance vs autonomy (↔ DP3); A4 engagement / interaction mode (↔ DP4); A5 disciplinary approach (↔ DP5).
- **Family B — beyond-framework attention (not scored by the PAI).** B1 monotony / repetitiveness of the sequence; B2 progression / pacing across activities; B3 holistic coherence of the plan; B4 prose quality / specificity of the descriptions; B5 personalization to the specific learner beyond the coded conditions; B6 affective and motivational fit.

**Procedure.** To validate the procedure, two educators independently coded a stratified 20% subsample of the weakest-aspect texts (n = 99), blind to the automated codes. Agreement is reported in Section 4.2, and only the codes it supports are used in the analysis. The full codebook with anchor terms and worked examples in both languages, the coding script, and the two coding sheets are released with the code.